# A Sub-grid Scale Energy Dissipation Rate Model for Large-eddy Spray Simulations

Hongjiang Li*, Christopher J. Rutland
Engine Research Center
University of Wisconsin-Madison
Madison, WI 53706
Francisco E. Hernández Pérez, Hong G. Im
Clean Combustion Research Center
King Abdullah University of Science and Technology
Thuwal 23955-6900, Saudi Arabia

**Abstract**
In high Reynolds number turbulent flows, energy dissipation refers to the process of energy transfer from kinetic energy to internal energy due to molecular viscosity. In large eddy simulation (LES) with one-equation turbulence models, the energy dissipation process is modeled by a rate term in the transport equation of the subgrid-scale (SGS) kinetic energy, $k_{sgs}$. Despite its important role in maintaining a proper energy balance between the resolved and SGS scales, modeling of the energy dissipation rate has received scarce attention. In this paper, a SGS model belonging to the dynamic structure family is developed based on findings from direct numerical simulation (DNS) studies of decaying isotropic turbulence. The model utilizes a Leonard-type term, a SGS viscosity, and a characteristic scaling term to predict the energy dissipation rate in LES. *A posteriori* tests of the model have been carried out under direct-injection gasoline and diesel engine-like conditions. Spray characteristics such as penetration rates and mixture fractions have been examined. It is found that the current SGS model accurately predicts vapor-phase penetrations across different mesh resolutions under both gasoline and diesel spray conditions, due to its correct scaling of SGS energy dissipation rate with the SGS kinetic energy and LES fitter width. In contrast, the classic model that is widely used in the literature predicts a scaling of energy dissipation rate upon mesh resolution, exhibiting a noticeable mesh dependence.

*Corresponding author: hongjiang.li@wisc.edu

**Introduction**

In turbulent flows at very high Reynolds number, kinetic energy is transferred from large eddies to small eddies via a cascade process, which proceeds further from these small eddies to even smaller ones [1]. This process continues until the molecular viscosity is effective in dissipating the kinetic energy. This notion is important because it puts the energy dissipation at the end of the energy cascade process.

In the context of large eddy simulation (LES) and subgrid-scale (SGS) modeling, depending on which turbulence model is used in the LES computation, the term "SGS energy dissipation rate" may have different meanings. For a zero-equation LES turbulence model, the "SGS energy dissipation rate" may refer to the energy flux from the resolved field to the SGS fields. For the one-equation LES turbulence model like the dynamic structure non-viscosity model developed by Pomraning and Rutland [2], the term refers to the energy removal from the SGS kinetic energy.

Research on modeling of the energy dissipation rate in LES of turbulent engine flows is very scarce. A classic model is written as [3]

$$\varepsilon_{sgs} = C\bar{\rho}\frac{k_{sgs}^{3/2}}{\Delta} \quad (1)$$

where $C, \rho, \Delta, k_{sgs}$ are a model constant, the gas-phase density, LES filter width, and SGS kinetic energy, respectively. The "overbar" represents the spatial filtering. The model was derived based on dimensional analysis, an approach commonly used in turbulence modeling. As Equation 1 demonstrates, this model guarantees positive dissipation (i.e., a sink for $k_{sgs}$), and thus better numerical stability. However, the value of $C$ requires *a priori* knowledge of the target flow and often needs to be tuned to provide better predictions [4].

For two-phase turbulent flows, the description of liquid particle – carrier phase turbulence interaction is more complicated and challenging in CFD simulations. Studies by Elghobashi [5, 6] show that the coupling between two-phases may be carrier phase → liquid particle (one-way), carrier phase ↔ liquid particle (dual-way), or carrier phase ↔ liquid particle ↔ liquid particle (four-way), depending on the volume fraction of particles. In turbulent engine flows, these coupling regimes may co-exist due to the complex spray structures. Bharadwaj and Rutland [7] studied the effect of spray-induced SGS kinetic energy on LES spray simulations and developed a spray source model to describe the liquid particle ↔ SGS carrier phase interactions. In their model, the SGS energy dissipation rate is still described by Equation 1.

Equation 1 also shows that the SGS energy dissipation rate scales with $k_{sgs}^{3/2}$. This power law scaling $\varepsilon_{sgs} \propto k_{sgs}^v$ has indeed been observed in various studies [4, 8]. The scaling factor, $v$, however, varies from 0.5 to 1.0 depending on the filter size and the flow conditions [8]. The *ad hoc* scaling of $\varepsilon_{sgs} \propto k_{sgs}^{3/2}$ in Equation 1 is therefore poorly justified. Recently, analogous Leonard-type SGS dissipation rate models have been developed [4]. *A priori* test results show that those models perform better than Equation 1. However, *a posteriori* testing in engine sprays has yet to be conducted to further evaluate their performance. Another approach to modeling the energy dissipation rate is to solve a transport equation for $\varepsilon_{sgs}$. One example is given by Pomraning and Rutland [2]. Nevertheless, this brings more difficulties such as unclosed terms in the $\varepsilon_{sgs}$-transport equation.

Recognizing the importance of the SGS energy dissipation rate in maintaining a proper energy balance between the resolved and SGS scales, and the fact that its modeling has received little attention, we re-examine the classic model and propose a model belonging to the dynamic structure family. In the following, the model is presented and applied to spray simulations at conditions relevant to engines, and concluding remarks are provided.

**LES Governing Equations**

LES gas-phase equations are obtained by applying spatial filtering to the fundamental conservation equations of mass, momentum, and energy. In the Lagrangian-Eulerian approach, fuel droplets are treated as point processes that occupy zero volumes. Contributions from droplets to the gas-phase conservation equations are then treated as source terms. A detailed description of the LES gas- and liquid-phase governing equations can be found in Refs. [7, 9]. For brevity, only the final form of the momentum equation is given

$$\frac{\partial \bar{\rho}\langle u_i\rangle}{\partial t} + \frac{\partial \bar{\rho}\langle u_i\rangle\langle u_j\rangle}{\partial x_j} = \frac{-\partial \bar{P}}{\partial x_i} + \frac{\partial \tau_{ij}}{\partial x_j} - \frac{\partial \bar{\rho}\Gamma_{ij}}{\partial x_j} + S_i \quad (2)$$

where $P, \tau_{ij}, \Gamma_{ij}, S_i$ are gas-phase pressure, molecular stress tensor, SGS stress tensor, and spray source term, respectively. The "brackets" represent the Favre filtering.

The SGS stress tensor, $\Gamma_{ij}$, cannot be directly computed and therefore a turbulence model is needed to close Equation 2. In practical applications such as internal combustion engine flows, the choice of model for $\Gamma_{ij}$ is of great importance. Thorough reviews on this subject have been given by Rutland [10] and Celik et al [11]. In this study, a mixed one-equation model developed by Tsang et al. [12] based on the work of Pomraning and Rutland [2] is used

$$\Gamma_{ij} = \frac{2L_{ij}}{L_{mm}}k_{sgs} - 2\nu_{noz}(\langle S_{ij}\rangle - \frac{1}{3}\delta_{ij}\langle S_{kk}\rangle) \quad (3)$$

where $L_{ij}$ is the Leonard stress tensor defined as $L_{ij} = \widehat{\langle u_i\rangle\langle u_j\rangle} - \widehat{\langle u_i\rangle}\widehat{\langle u_j\rangle}$, $\nu_{noz}$ is an artificial viscosity in the near nozzle area, and $S_{ij}$ is the strain rate tensor. The operator $\widehat{\phantom{xx}}$ denotes a test level filtering on an additional cell layer around the computational cell.

The SGS kinetic energy, $k_{sgs}$, is determined using a transport equation given by Bharadwaj and Rutland [13]. The modeled form of this equation is written as

$$\frac{\partial \bar{\rho} k_{sgs}}{\partial t} + \frac{\partial \bar{\rho} k_{sgs}\langle u_j\rangle}{\partial x_j}$$
$$= \frac{\partial}{\partial x_j}\left(\mu_T \frac{\partial k_{sgs}}{\partial x_j}\right) - \bar{\rho}\Gamma_{ij}\frac{\partial\langle u_i\rangle}{\partial x_j} - \varepsilon_{sgs} \quad (4)$$
$$+ \dot{W}_{s,sgs}$$

where $\mu_T$, $\dot{W}_{s,sgs}$ are the SGS turbulent viscosity and spray source term given in Ref. [9].

**Model Development**

The model parameter $C$ in Equation 1 often needs to be tuned across different mesh resolutions to compensate for the poor scaling of $\varepsilon_{sgs}$ with the SGS kinetic energy, $k_{sgs}$, and LES filter width, $\Delta$. In the authors' experience, its value also needs to be scaled with a characteristic length in order to reach good mesh independence of the vapor-phase penetration. In this work, the characteristic length is selected to be the LES filter size in the near-nozzle exit, $\Delta_{noz}$. Once $C$ is tuned for mesh $a$ under specific conditions, its value for mesh $b$ is determined by

$$C_b = C_a \frac{\Delta_{noz,b}}{\Delta_{noz,a}} \quad (5)$$

Another formulation of $\varepsilon_{sgs}$ is developed based on the work of Chumakov and Rutland [4] and Pomraning and Rutland [2]. The proposed form of this model is given by Chumakov and Rutland [4]

$$\varepsilon_{sgs} \approx \mu_C F\left[\frac{\partial\widehat{\langle u_l\rangle}}{\partial x_j}\frac{\partial\langle u_l\rangle}{\partial x_j} - \frac{\widehat{\partial\langle u_l\rangle}}{\partial x_j}\frac{\widehat{\partial\langle u_l\rangle}}{\partial x_j}\right] \quad (6)$$

where $\mu_C$ is a characteristic viscosity and $F$ is a function determined from *a priori* tests. The remaining part inside the bracket is a Leonard-type term, which is always positive for non-negative filters [3]. Examples of such filter are the box, Gaussian, and linear filters [4]. The Fourier cut-off filter, on the other hand, assumes negative values in real space so that the Leonard-type term may be negative as well. However, it is rarely used in engineering applications. Therefore, in this work a positive $\varepsilon_{sgs}$ (i.e., sink for $k_{sgs}$) is always implied.

Starting from the filtered momentum equation, a natural approach to estimate $\mu_C$ would be using the molecular viscosity, $\mu$. However, *a posteriori* test of this approach shows that this approximation would lead to severe underestimation of the energy dissipation rate term, $\varepsilon_{sgs}$, hence a proper budget of the SGS kinetic energy, $k_{sgs}$, is not achieved. Another approach, which is considered here, consists in replacing it with the SGS viscosity: $\mu_C = \mu_{SGS}$. In this study, the formula proposed by Lilly [14] is used

$$\mu_{sgs} = C_k\bar{\rho}\Delta k_{sgs}^{1/2} \quad (7)$$

Finally, the shape of the function $F$ needs to be found. From the study of Chumakov and Rutland [8], an appropriate approximation can be written as

$$F = C_\epsilon k_{sgs}^\alpha \Delta^\beta \quad (8)$$

where α and β are scaling factors determined from *a priori* tests. Chumakov [8] found through DNS of forced isotropic turbulence that $\varepsilon_{sgs} \propto k_{sgs}^{1/2}$ for $\Delta$ close to the forcing scale. For $\Delta$ in the near viscous range, $\varepsilon_{sgs} \propto k_{sgs}^{1.0}$. Since the focus of this work is on engine sprays, where the size of $\Delta$ is considerably larger than the viscous length scale (i.e., Kolmogorov length scale), the value of α is set to be zero, which effectively leads to the following scaling: $\varepsilon_{sgs} \propto k_{sgs}^{1/2}$.

According to the classic model of energy dissipation rate (i.e. Equation 1), $\varepsilon_{sgs}$ scales inversely with $\Delta$. Chumakov [8] argues that in the near viscous range, $\varepsilon_{sgs}$ scales with the square of $\Delta$. Assuming this scaling also holds for $\Delta$ in the forcing range, we can argue that

$$\varepsilon_{sgs}$$
$$\approx \bar{\rho} C_k C_\epsilon \Delta^2 k_{sgs}^{1/2}\left[\frac{\partial\widehat{\langle u_l\rangle}}{\partial x_j}\frac{\partial\langle u_l\rangle}{\partial x_j} - \frac{\widehat{\partial\langle u_l\rangle}}{\partial x_j}\frac{\widehat{\partial\langle u_l\rangle}}{\partial x_j}\right] \quad (9)$$

The model constants $C_k$ and $C_\epsilon$ can be combined together for practical implementations. Following a simple dimensional check, we assume

$$C_k C_\epsilon = \frac{C_{sgs}}{d_{noz}} \quad (10)$$

The final form of $\varepsilon_{sgs}$ can therefore be written as

$$\varepsilon_{sgs} \approx \bar{\rho} \frac{C_{sgs}}{d_{noz}} \Delta^2 k_{sgs}^{1/2} \left[ \widehat{\frac{\partial \langle u_\iota \rangle}{\partial x_J} \frac{\partial \langle u_\iota \rangle}{\partial x_J}} - \frac{\widehat{\partial \langle u_\iota \rangle}}{\partial x_J} \frac{\widehat{\partial \langle u_\iota \rangle}}{\partial x_J} \right] \quad (11)$$

where the default value of $C_{sgs}$ is set to be 0.11.

**Experimental Conditions and Numerical Setup**

*A posteriori* testing of the SGS energy dissipation rate model (i.e., Equation 10) is carried out under both direct injection gasoline and diesel engine-like conditions, which are selected from the engine combustion network (ECN). The gasoline case, termed as "spray G", corresponds to a non-reacting early injection case in direct injection spark-ignited (DISI) engines [15]. The remaining one, termed as "spray H", represents combustion conditions relevant to diesel engines. The injection conditions and injector parameters of both spray experiments are given in Table 1.

The OpenFOAM CFD package was used for implementing the models and running LES simulations [16]. For spray G, simulations based on the classic model (i.e., Equation 1) with and without scaling in Equation 5 are also conducted for comparison. Simulations were run using a cubical domain with a length of 100 mm. CFD meshes with minimum cell sizes of 1.0, 0.5, and 0.375 mm were modified to include a cutout approximating the shape and location of the DISI multi-hole injector [17]. The integrated atomization/breakup model and the SGS dispersion model developed by the authors are used to describe the spray breakup and turbulent dispersion processes [9]. The CFD mesh used for spray H simulations has a dimension of 80 × 80 × 100 mm with static mesh refinement to cover the liquid and vapor region. Detailed descriptions of the simulation setup can be found in Reference [9].

**Table 1.** Conditions and injector specifications for ECN evaporating spray H and spray G.

| Data type | Spray H | Spray G |
|---|---|---|
| Fuel | n-heptane | iso-octane |
| Ambient temperature (K) | 1000.0 | 573.0 |
| Ambient pressure (MPa) | 4.33 | 5.97×10⁻¹ |
| Ambient velocity (m/s) | ≈ 0 | ≈ 0 |
| Fuel temperature (K) | 373 | 363 |
| Injector orifice diameter (mm) | 0.100 | 0.165 |
| Injection pressure (MPa) | 154.33 | 20.0 |
| Injection duration (ms) | 6.8 | 0.78 |
| Injected fuel mass (mg) | 17.8 | 10.0 |

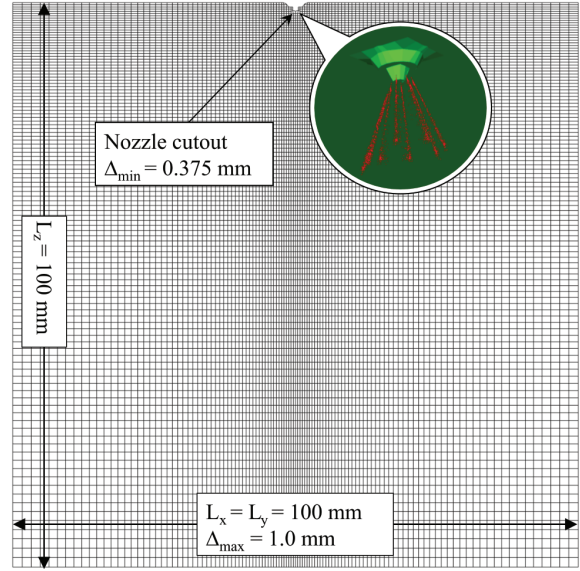

**Figure 1.** x-y cut plane of the cubical mesh for spray G simulations. Note that there is a nozzle cutout approximating the shape and location of the injector to guide the near-nozzle flows [17].

**Results and Discussion**

Results corresponding to the ECN spray G conditions will be discussed first. Subsequently, results corresponding to the evaporating spray H conditions will be presented.

**Evaporating Spray G**

Figure 2 presents the vapor-phase penetrations from experiments and simulations for spray G. Experimental data are taken from measurements by Sandia National Laboratory (SNL) and Institute Motori (Ist Motori) [15]. Simulation results are from single LES realizations only. The results at the top, middle, and bottom plots correspond to the classic model, the classic model with scaling, and the dynamic SGS model, respectively. Starting with the top and middle plots, it can be seen that while the scaling of the model parameter reduces the mesh sensitivity, the variance among all three meshes is still noticeable. The classic model severely over-predicts the vapor-phase penetration with the 0.375 mm mesh. One possible explanation is that the classic model over-estimates the energy dissipation rate with the finest mesh. This hypothesis will be explored in detail later. On the other hand, results for the proposed dynamic SGS model show significantly improved mesh independence. The vapor-phase penetrations after the end of injection also show improved matching with the SNL data.

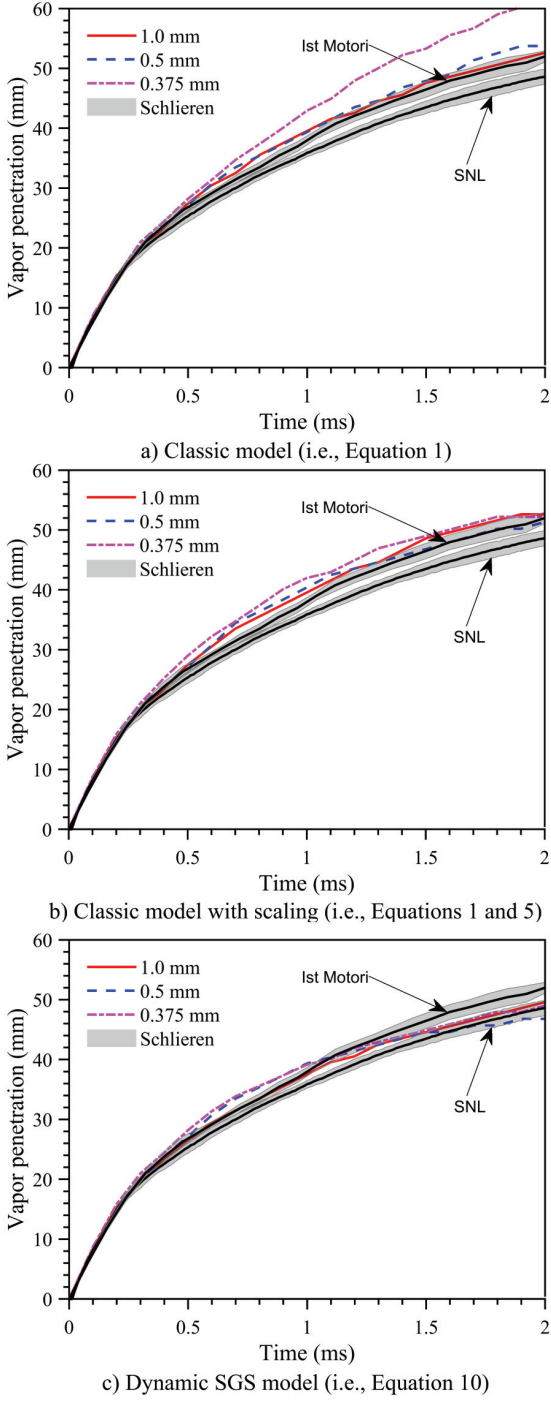

**Figure 2.** Vapor-phase penetrations for spray G. LES simulations were ran using three meshes with minimum cell sizes of 1.0, 0.5, and 0.375 mm. Experimental measurements were performed by Sandia National Lab (SNL) and Institute Motori (Ist Motori) [15]. The predicted results are processed with a local fuel mass fraction of 0.001.

To help explain the different mesh dependencies observed in Figure 2, the following discussion will be focused on the energy dissipation rate, $\varepsilon_{sgs}$, for the 0.375 mm mesh. Results on the x-y plane are plotted in Figure 3 at three selected times of 0.3, 0.7, and 1.1 ms ASOI. Starting with the results at 0.3 ms ASOI, presented on the left column, an immediate observation is that the predicted shapes and characteristic sizes of the $\varepsilon_{sgs}$ distribution are very similar among all three cases. However, Figure 3a shows higher values of $\varepsilon_{sgs}$ in the near-nozzle area compared to the other two cases. Recall that the time derivative of the SGS kinetic energy, $k_{sgs}$, is directly linked to $\varepsilon_{sgs}$ (see Equation 4)

$$\frac{\partial k_{sgs}}{\partial t} \propto -\varepsilon_{sgs} \qquad (12)$$

This implies that too much $k_{sgs}$ may have been removed from the SGS field with the classic model. Proceeding to the middle column of Figure 3, where results are shown for 0.7 ms ASOI, noticeable differences among all three models can be observed. Compared to the other two models, the classic model continues to show higher $\varepsilon_{sgs}$ in the near-nozzle area, but the characteristic footprint below the injector is much smaller. This suggests that the energy budget of $k_{sgs}$ and the energy cascade may not be properly described with the classic model. The differences become even larger at 1.1 ms ASOI, as the right column in Figure 3 shows. Figure 3b and 3c exhibit a plume collapsing toward the injector centerline, which correlates well with the findings in Ref. [18]. Figure 3a, on the other hand, does not show such behavior as evidenced by the distinct boundaries between the plume pairs. Note that Figure 3c also shows more variance on the outer side, which may be introduced by the Leonard-type term in Equation 11.

Continue focusing on the energy cascade, one can now examine the ratio of $k_{sgs}$ to the total kinetic energy, calculated as the summation of $k_{sgs}$ and resolved kinetic energy, K. Results are plotted in Figure 4. Similar findings to $\varepsilon_{sgs}$ are observed for the kinetic energy ratio. According to Equation 12, the over-estimation of $\varepsilon_{sgs}$ with the classic model will give under-estimated $k_{sgs}$, which is supported by the smaller kinetic energy ratios found in Figure 4a. The energy removal from the resolved field is also under-estimated, since, in the dynamic structure turbulence closure, the SGS stress tensor scales with $k_{sgs}$ (see Equation 3). As a consequence, the vapor-phase penetration is over-predicted as shown in Figure 2a.

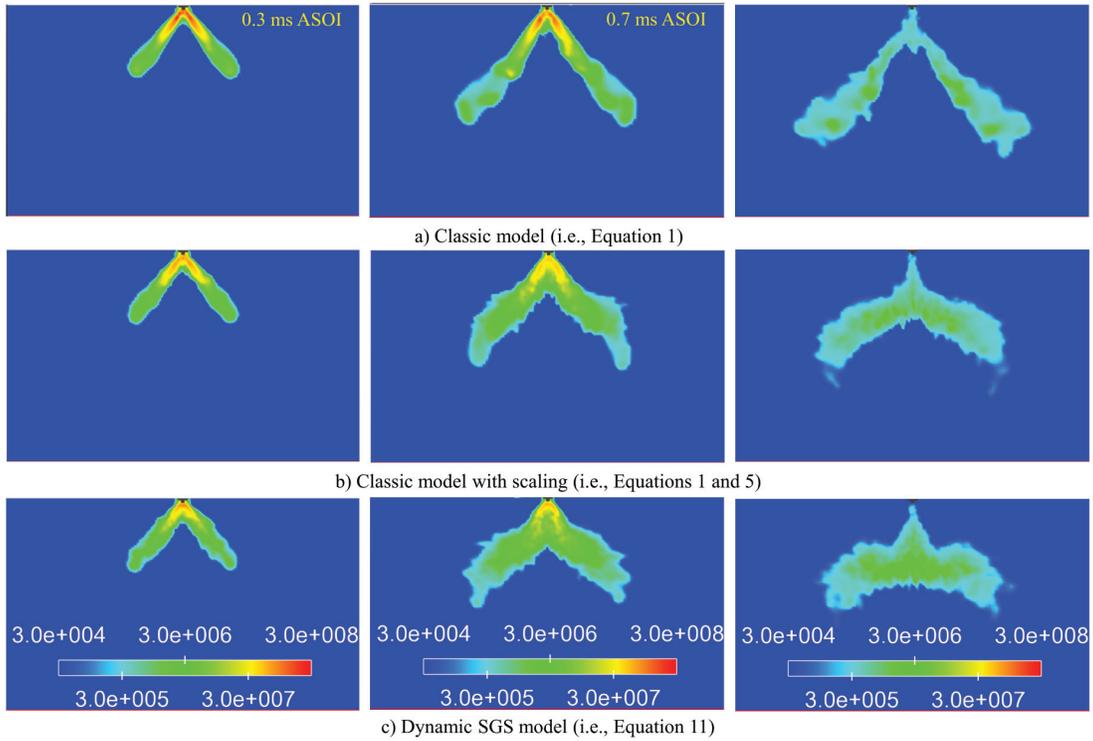

**Figure 3.** SGS energy dissipation rates with 0.375 mm mesh. Results are plotted on the x-y plane through nozzle centerline. From left to the right, results are presented at 0.3, 0.7, and 1.1 ms ASOI, respectively.

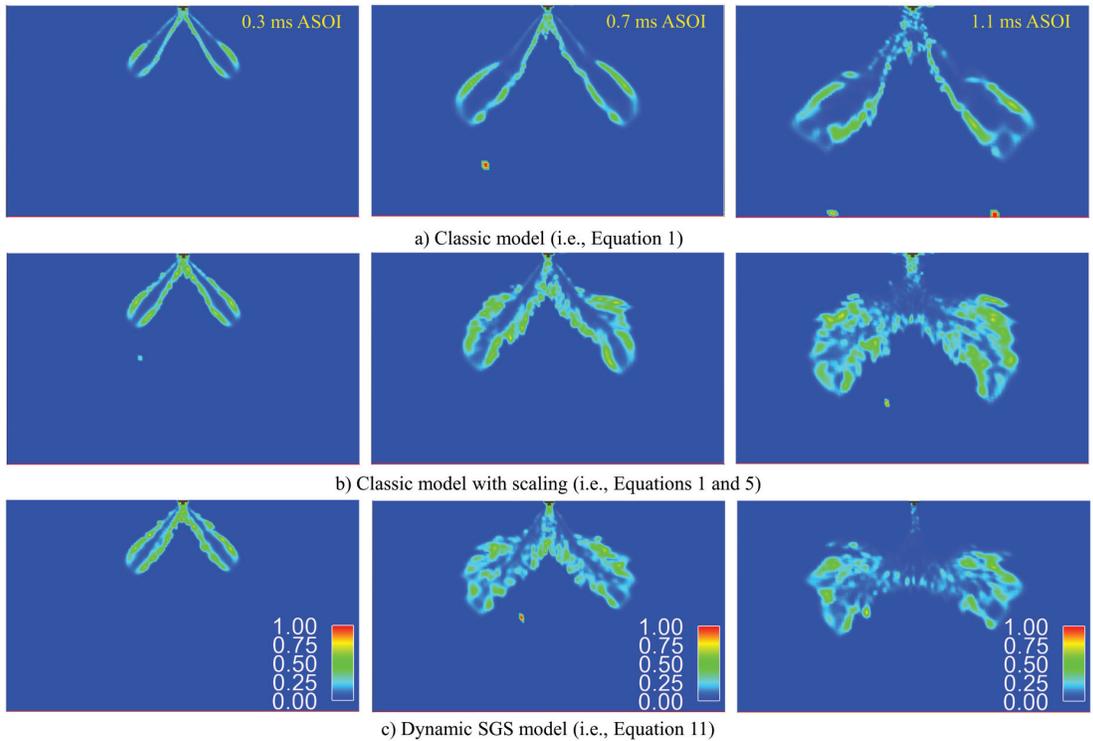

**Figure 4.** The ratio of SGS kinetic energy to total kinetic energy with 0.375 mm mesh. Results are plotted on the x-y plane through nozzle centerline. From left to the right, results are presented at 0.3, 0.7, and 1.1 ms ASOI, respectively.

**Diesel Spray H**

The proposed dynamic SGS model is then applied to the evaporating, non-reacting spray H case for further evaluation. The characteristic penetrations predicted by the LES simulations with three mesh resolutions of 0.5, 0.375, and 0.25 mm are plotted along with experimental data in Figure 5. Note that a different value of the model parameter $C_{sgs}$ is employed (0.18 instead of the default value of 0.11) for a better matching with the experimental data. It can be seen that all three simulations over-predict the vapor-phase penetration length shortly after the start of injection, with worse agreement yielded with finer mesh. The predicted fuel vapor seems to travel downstream with a constant velocity until 0.2 ms ASOI, while the measured one becomes parabolic around 0.1 ms ASOI. This discrepancy leads to the over-prediction of fuel vapor until approximately 0.5 ms ASOI. The reason of this behavior is unknow at this point. Overall, the model gives fairly satisfactory results, especially after 0.5 ms ASOI, across all three mesh resolutions.

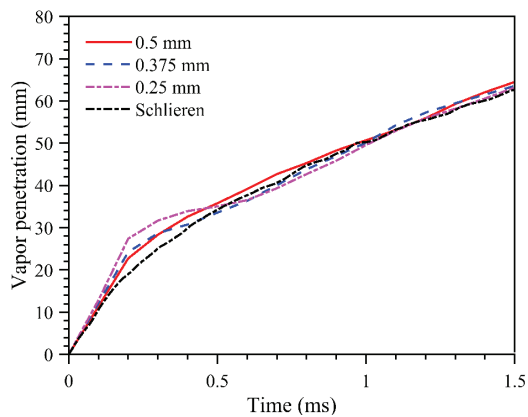

**Figure 5**. Vapor-phase penetration results for spray H case. Simulation results are obtained from single-LES realizations only. Experimental data are taken from Schlieren measurements conducted by SNL [19].

Quantitative mixing information was obtained at 0.9 ms ASOI. The radial profiles of the fuel vapor mixture fraction at 20 and 40 mm below the injector exit are plotted in Figures 6 and 7, respectively. In each figure, the experimental data from Rayleigh scattering measurements conducted by SNL [19] are plotted with solid lines representing the mean and shaded areas representing the standard deviations. Starting with Figure 6, in which the vapor mixture fractions resemble normal distributions, a relatively good agreement can be seen between experiments and LES simulations except near the nozzle centerline. Moving further downstream at 40 mm axial position, both predicted and measured vapor mixture fraction profiles show more fluctuations as shown in Figure 7. This is expected since the spray-induced turbulent flow becomes more developed in the downstream. With the finest mesh of 0.25 mm cell size, the dynamic SGS model over-predicts the mixture fraction near the injector centerline in both figures, but the agreement becomes better at locations away from injector centerline.

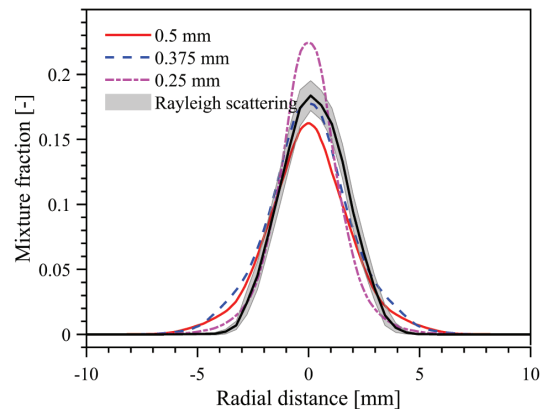

**Figure 6**. Radial distribution of the fuel vapor mixture fraction at 20 mm axial position for the evaporating, non-reacting spray H condition of 1000 K. Experimental data are plotted with a shaded area representing the standard deviation.

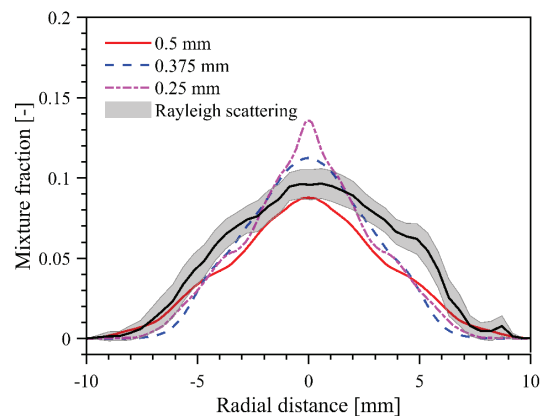

**Figure 7**. Radial distribution of the fuel vapor mixture fraction at 40 mm axial position for the evaporating, non-reacting spray H condition of 1000 K. Experimental data are plotted with a shaded area representing the standard deviation.

Figure 8 shows the vapor-phase envelops at selected times of 0.5, 1.0, and 1.5 ms ASOI. Each row represents a different mesh resolution. In all plots, vapor-phase envelops identified in the Schlieren measurements are overlaid as solid red lines. As expected, early in time (0.5 ms ASOI), the vapor-phase shows little difference among simulations. As time progresses, however, it is apparent that as the mesh becomes finer, the narrower vapor envelope persists. Additionally, the fuel mixture fraction on the injector centerline shows higher values in the downstream for the finer mesh. This trend corresponds to radial distributions shown in Figures 6 and 7.

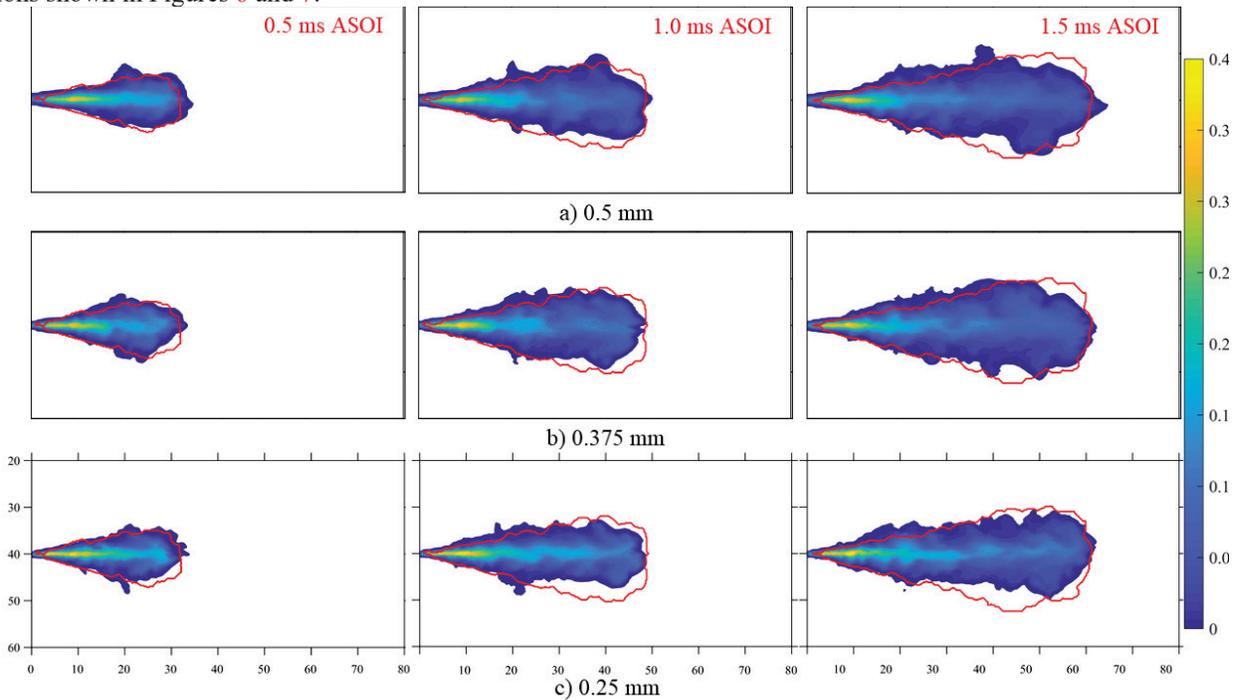

**Figure 8**. Contour plots of the fuel mixture fraction at various selected times ASOI for three meshes with minimum cell sizes of 0.5, 0.375, and 0.25 mm. The vapor-phase envelops identified in the Schlieren measurements are plotted as solid red lines on each plot. The predicted results are taken from single realizations only.

**Summary and Conclusions**

A sub-grid scale energy dissipation rate model is presented in this paper. The model belongs to the dynamic structure SGS model family, in which the modeled energy dissipation rate term is constructed from the corresponding Leonard term and a scaling term, which was developed based on the findings from reported DNS studies of decaying isotropic turbulent flows. *A posteriori* tests have been carried out under both direct injection gasoline and diesel engine-like conditions, and the LES results have been compared against experimental data available in the literature.

It has been found that the SGS model achieves good mesh independence among various grids utilized in this paper. This is attributed to the improved scaling of the energy dissipation rates with the SGS kinetic energy and LES filter width. The classic model, which was developed on dimensional analysis grounds, cannot guarantee accurate budgets of SGS kinetic energy across different mesh resolutions without tuning the model parameter for each mesh.

*A priori* testing of the SGS model will be carried out in the near future, including a comparison to DNS data from simulations of decaying isotropic turbulent flows.

**Acknowledgments**

This work is supported by the FUELCOM collaborative project between King Abdullah University of Science and Technology and Saudi Aramco Oil Co. The computing resources at UW-Madison High Throughput Computing were used to obtain results presented in this publication.